\documentclass[a4paper,11pt]{article}
\pdfoutput=1 

\usepackage{jheppub} 

\usepackage[T1]{fontenc} 

\newcommand{\tof}{t_\textsc{\scriptsize of}}

\usepackage{xcolor}

\title{\boldmath Identification of Long-lived Charged Particles using Time-Of-Flight Systems at the  Upgraded LHC detectors}

\author[a,1]{O. Cerri,\note{Corresponding author.}}
\author[a]{S. Xie}
\author[a, b]{C. Pena}
\author[a]{and M. Spiropulu}

\affiliation[a]{California Institute of Technology, Pasadena, CA 91125, U.S.A.}
\affiliation[b]{Fermi National Accelerator Laboratory, Batavia, Illinois}

\emailAdd{ocerri@caltech.edu}
\emailAdd{sixie@caltech.edu}
\emailAdd{cristian.pena@caltech.edu}
\emailAdd{smaria@caltech.edu}

\abstract{
We study the impact of precision timing detection systems on the LHC experiments' long-lived particle search program during the HL-LHC era. We develop algorithms that allow us to reconstruct the mass of such charged particles and perform particle identification using the time-of-flight measurement. We investigate the reach for benchmark scenarios as a function of the timing resolution, and find sensitivity improvement of up to a factor of ten over searches that use ionization energy loss information, depending on the particle's mass.

}

\begin{document} 
\maketitle
\flushbottom

\section{Introduction}
\label{Intro}
The CMS and ATLAS collaborations have been exploring precision timing detector concepts intended to enable the time-of-arrival measurement of charged particles with a resolution of a few tens of picoseconds~\cite{MTDTechProposal,Allaire:2302827}. Such a detector promises to significantly mitigate the impact of the large number of simultaneous collisions within the same bunch crossing (pileup) expected for the High-Luminosity LHC. 

The proposed HL-LHC beamspot extends about $10$~cm along the beam-axis and about $200$~ps in time. For the Phase 2 CMS tracker, the rate of spurious merging of vertices begins to become significant for vertices separated by less than one mm. For an average number of collisions per bunch-crossing of $140$--$200$, it has been shown that the fraction of spurious tracks from pileup collisions
associated with the reconstructed primary vertex is between $20$ and $30\%$. With the addition of a time measurement for tracks, this fraction of spurious track-to-vertex association is reduced to about $5\%$~\cite{MTDTechProposal}. As a result, significant improvements on the efficiency for particle identification, including isolated leptons and photons, forward jet identification, as well as missing transverse energy resolution recovery, are expected. 

Previous studies have demonstrated that significant reach enhancement of the HL-LHC physics program can be realized by using a combination of the time-of-arrival measurement with secondary vertices to reconstruct the mass of long-lived exotic neutral particles~\cite{MTDTechProposal} produced by proton-proton collisions at the LHC. In this paper, we complement those studies by enabling the reconstruction of the masses of long-lived exotic charged particles or heavy stable charged particles (HSCP) by using the position of the production vertex, the time-of-flight, and the track momentum. The resonance mass reconstruction yields a uniquely enhanced capability to identify new particles and to suppress backgrounds for searches of long-lived exotic particles. We discuss methods for reconstructing the time of the primary collision vertex and strategies for using the resulting time-of-arrival measurement to infer the mass of the HSCP. We demonstrate the effectiveness of these methods by evaluating the vertex time resolution and the mass resolution, and we show that an improvement in search sensitivity of a factor of $5$--$10$ can be achieved for HSCP masses above $300$~GeV.  

The paper is organized as follows. We discuss the details of the time-of-flight particle identification in Section~\ref{sec:TOFPID}, the benchmark signal model and Monte Carlo simulation in Section~\ref{sec:simulation}, the time-of-flight (TOF) and resonance mass reconstruction in Section~\ref{sec:TOFRECO}, and the analysis and search sensitivity estimate in Section~\ref{sec:benchmark_analysis}. We conclude and summarize in Section~\ref{sec:summary}.

\section{Time-of-Flight Particle Identification}
\label{sec:TOFPID}
The time-of-flight $\tof$ refers to the time needed for a particle to travel between two spatial points. When the length of the particle trajectory ($L$) is known, one can compute the velocity of the particle as $\displaystyle \beta=L /c\tof$. Combining the latter with the momentum ($\vec{p}$) measurement of the particle, typically obtained by precision tracking detectors at colliders, it is possible to extract the particle mass via the relation
\begin{equation*}
\beta = \frac{p}{\sqrt{p^2 + m^2}}.
\end{equation*}
Using TOF measured to a resolution of a few tens of picoseconds
and detectors separated by about one meter from the collision point, one can obtain significant discrimination power between different mass hypotheses. Therefore, TOF measurements are a powerful tool for particle identification (PID). We illustrate its use for PID with a simplified example using the geometry of the CMS detector. The CMS detector is cylindrical and we define the $\hat{z}$ axis to be the line going parallel through the center of the cylinder. The total length of the cylindrical detector is equal to $H_d = 6 \text{ m}$. A time-of-arrival measurement is performed in a cylindrical detector layer located at a radius of $R_d = 1.29\text{ m}$ and the TOF is extracted using the time of the particle production vertex. The axial magnetic field is $B_z = 3.8 \text{ T}$. A particle is characterized by its transverse momentum $p_{T}$, its mass $m$ and its pseudo-rapidity $\eta = \text{-} \ln \tan \frac{\theta}{2}$, where $\theta$ is the polar angle measured from the $\hat{z}$ axis. The TOF measurement is assumed to have Gaussian uncertainty with an instrumental resolution of $\sigma_{\textsc{tof}}$ and the transverse momentum is measured with a resolution of a few percent. 

\begin{figure}[tbp]
\centering
\includegraphics[width=0.99\textwidth]{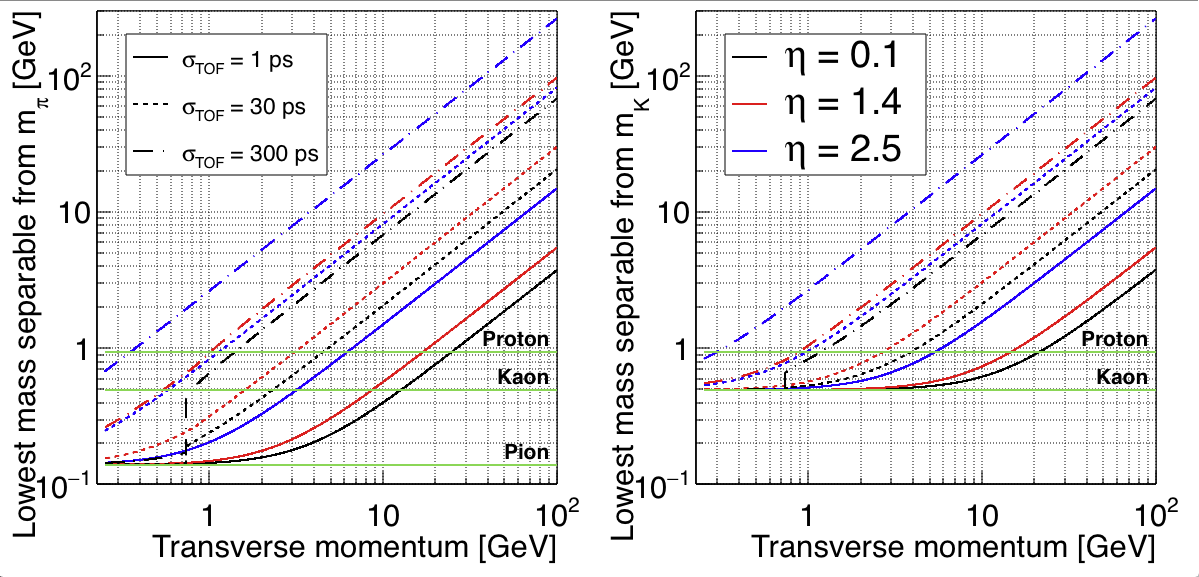}
\caption{Mass of the lighter particle that has at least a $3\sigma$ discrepancy in $\tof$ from the hypothesis $m = m_\pi$ (left) and $m = m_K$ (right) as a function of the particle transverse momentum. Different colors represents different pseudo-rapidity regions and different styles represent different time resolution scenarios as indicated in the legend.}
\label{fig:mass_vs_pt}
\end{figure}

To estimate the separation power between different mass hypotheses, we compute the mass for which we can achieve separation significance higher than $3\sigma$ ( $\text{p-val} < 0.03$ ) from the pion mass ($m_\pi$) hypothesis.
The TOF can be expressed, as a function of the particle kinematic variables as:
\begin{equation}
\label{eqn:TOF}
\tof = \frac{L(p_T, \eta)}{c p} \sqrt{p^2 + m^2}.
\end{equation}
Under the correct mass hypothesis $m = m^*$, 
\[
\tof^{(\text{meas})} - \tof|_{m = m^*} \sim N(0, \sigma_{\textsc{tof}})
\]
where $\tof^{(\text{meas})}$ is the measured value, $\tof|_{m = m^*}$ is the expected value for a mass equal to $m^*$ and $N(\mu, \sigma)$ is a Gaussian distribution with mean $\mu$ and standard deviation $\sigma$. Therefore, all the particles satisfying the relation
\[
\frac{L}{c \sigma_{\textsc{tof}}}\left|\sqrt{1+\frac{m^2}{p^2}} - \sqrt{1+\frac{m_\pi^2}{p^2}}\right| > 3
\]
have a $3-\sigma$ significance separation from the pion hypothesis.
In Figure~\ref{fig:mass_vs_pt} we show as a function of the particle $p_T$, the minimum mass that a particle must have to be incompatible with the charged pion (left) and kaon (right) hypothesis, as well as its evolution with $\eta$ and the TOF instrumental resolution $\sigma_{\textsc{tof}}$.
Three TOF resolution scenarios are considered: 300 ps, corresponding to the best time resolution currently available in CMS for calorimeter deposits; 30 ps, being an ambitious but reasonable target for the performance during HL-LHC operation; and 1 ps, with ambitious future technology. The considered pseudo-rapidity values correspond to tracks ending in the central barrel (0.1), near the intersection between barrel and endcap (1.4), and close to the forward limit of the detector acceptance (2.5).
In the HL-LHC scenario, with the expected CMS detector performance, separation between pions and kaons is  achievable up to about $5$-$7$~GeV in transverse momentum.

\begin{figure}[tbp]
\centering
\includegraphics[width=0.8\textwidth]{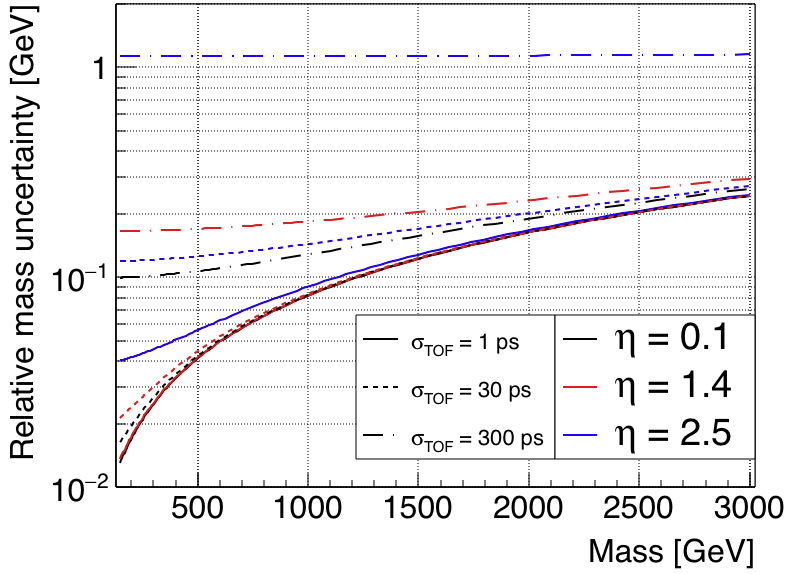}
\caption{Expected mass resolution as a function of mass reconstructed based on the TOF measurement as indicated in the legend, for stable charged particles with a transverse momentum equal to their mass.}
\label{fig:mass_uncertainty}
\end{figure}

Similarly, we can use the TOF measurement to reconstruct the mass resonance of an exotic massive particle. For pair-produced particles at the LHC, its $p_T$ is typically of the same order as its mass. With this assumption, the mass of the particle estimated through Eqn~\ref{eqn:TOF} using the measured TOF will exhibit a peaked shape reflecting the shape of the particle resonance convoluted with detector resolution effects. This peak structure is a powerful discriminator between a new particle and the background; as the peak becomes narrower, the discrimination power increases. The resolution of the mass resonance can be expressed as: 
\begin{equation}
\label{eqn:massResolution}
(\Delta m)^2 = m^2\left[\left(\frac{\Delta p_T}{p_T}\right)^2 + \left(\frac{1}{1-\beta^2}\right)^2\left(\frac{\sigma_{\textsc{tof}}}{\tof}\right)^2\right]
\end{equation}
In Figure~\ref{fig:mass_uncertainty} we show the expected relative mass resolution as a function of the particle mass, as well as its evolution with $\eta$ and the TOF instrumental resolution $\sigma_{\textsc{tof}}$. In the HL-LHC scenario, with the expected CMS detector performance, the estimated resolution on the mass of a 1 TeV particle is about $7\%$ .

\section{Signal Model and Monte Carlo Simulation}
\label{sec:simulation}
Many extensions of the standard model (SM) include heavy, long-lived, charged particles \cite{BAUER2010280,FAIRBAIRN20071,DREES1990695} that might be produced at the LHC with a speed significantly less than the speed of light. Those with lifetimes greater than a few nanoseconds can travel distances larger than the typical LHC detector and appear stable. These particles are generically referred to as heavy stable charged particles (HSCPs). Precise time measurements can  improve the detection of such particles.
We consider signals consisting of HSCPs that interact via the strong force and hadronize with SM quarks to form R-hadrons. Similar to reference~\cite{CMS-PAS-EXO-16-036}, we consider the pair production of top squarks $\tilde{t}_1$, with masses in the range 100-2500 GeV, generated under the Split SUSY scenario. For the purpose of this study all other masses of the SUSY spectrum are sequestered and considered to be larger than O(10) TeV.

As there are no SM processes that produce a pair of HSCPs, the dominant background at the LHC is QCD multijet production, which can mimic the signature of the HSCPs due to instrumental resolution. The instrumental resolution is independent of the specific production process and QCD multijet production is dominant because of its large cross section.

For all the Monte Carlo simulation samples, \textsc{Pythia} v8.230~\cite{sjostrand2015introduction}, with the 4C tune \cite{corke2011interleaved}, is used to generate proton proton collisions at 14 TeV. Modeling of multiple interactions and initial and final state radiation are turned on.
The signal simulation sample is generated, accounting for all tree-level squark pair-production processes, including gluon-gluon, q-$\bar{\rm{q}}$ and q-q initial states. Hadronization to form R-hadrons is activated and all possible R-hadrons are allowed based on the initialized squark masses and the constituent masses of the other partons in the hadron.
QCD multijet background samples are generated in the following bins of $\hat{p}_t$, defined as the $p_{T}$ of the first two partons: 30-50, 50-100, 100-150, 150-185, 185-300, 300-600 and 600-infinity (values expressed in GeV).
The pile-up collisions are generated using soft QCD processes which are then mixed with background or signal interaction.

Detector effects are simulated using the Delphes~3.4.1~\cite{deFavereau:2013fsa} parametric simulation with a dedicated configuration card used to emulate the performance of the CMS detector after the HL-LHC upgrade. Events are simulated for a scenario with an average pile-up of 140 and the beam-spot is assumed to be Gaussian with $\sigma$ of $160$~ps and $5.3$~cm for the time and z coordinates respectively. A negligible beam-spot size in the $x-y$ plane has been assumed, following the operational parameters of the HL-LHC \cite{Apollinari:2120673}.
The tracker performance is parametrized for the geometry of the CMS HL-LHC upgrade design~\cite{mersi2012software}. The timing detector is simulated with Gaussian time-of-arrival resolution with $\sigma$ of $30$~ps for all the tracks reaching $R=1.29 \text{ m}$ with $|\eta| < 3$. This parametrization is consistent with testbeam results~\cite{apresyan2016test,APRESYAN2018158} and the CMS MIP timing detector technical proposal~\cite{MTDTechProposal}. We have developed new Delphes modules to implement the mass reconstruction and PID based on the TOF measurement of tracks. A new module to reconstruct vertices simultaneously using both  the space and time measurements of tracks was  developed, as well as a module to implement the TOF reconstruction. Details of these reconstruction modules are described in the following Section (Section ~\ref{sec:TOFRECO}).

\section{Space-time Vertex and TOF Reconstruction }
\label{sec:TOFRECO}

To reconstruct the velocity of a particle, it is necessary to measure the time difference and length between two points along its trajectory. The tracker allows a measurement of the track length with a typical relative resolution of order $10^{-3}$. 
However, only the absolute time-of-arrival of the particle at the point where the particle trajectory intersects the timing detector layer is measured. In order to measure the TOF between the particle production vertex (inner point) and the impact point on the timing detector layer (outer point), the time of the collision that produced the charged particle must be obtained. The naive approach of using the bunch-crossing time as the reference time will yield a resolution not better than $160$~ps, due to the spread in time of the collision beam-spot. Instead, we developed a method that precisely reconstructs the space-time coordinate of all vertices. This allows a velocity measurement to be made for all tracks which can be associated to a vertex. Once associated to a vertex, the vertex time is considered as the production time of the particle and it is used to compute the TOF.

\subsection{Space-time Vertex Reconstruction}
\label{sec:vertexing}
Each track reconstructed in the collision event is clustered together using a deterministic annealing (DA) algorithm~\cite{rose1998deterministic} to reconstruct collision vertexes. DA algorithms have been shown to perform well for vertex finding in CMS  using the three dimensional spatial coordinates~\cite{chabanat2005deterministic}.
We have developed a new Delphes module that implements the DA algorithm, extending it to include the time-coordinate. 
Beside the specific interest for the TOF measurement, the use of the time coordinate in the vertex reconstruction procedure to distinguish tracks from collisions that are very closely separated in space will be crucial at the HL-LHC to maintain the current level of detector performance~\cite{MTDTechProposal}, due to the large amount of pile-up expected.

As shown in \cite{fruehwirth2003new}, for clustering it is convenient to substitute tracks with representative points. We substituted each track with the time ($t_\textsc{ca}$) and the position ($z_\textsc{ca}$) of closest approach to the beam axis. The position of closest approach $z_\textsc{ca}$ is extracted as one of the parameters in the track fit, while the time of closest approach is computed, under a given mass hypothesis $m$, with the following relation:
\[
t_\textsc{ca} = t_\textsc{td} - \int_T \frac{\hat{\beta}}{c\beta}d\vec{x} = t_\textsc{td} - \frac{L}{c}\frac{\sqrt{p^2 + m^2}}{p}
\]
where the line integral is computed along the track trajectory ($T$) from the point of closest approach to the location of the Timing Detector (TD), and $t_\textsc{td}$ is the measured time-of-arrival at the TD. 
To associate each track with a representative point, it is assumed that all the tracks are from charged pions with mass $m_{\pi}$. This is a good first order approximation, as the majority of charged particles produced at the LHC are pions or have masses close to the pion mass. For tracks from particles that have significantly different mass, $t_\textsc{ca}$ will be shifted with a magnitude that depends on the particle's momentum and $\eta$. This shift will either result in the track being unsuccessfully clustered, or result in the track being clustered to the wrong vertex. The signal model we are considering in this work produces prompt heavy long-lived particles, so the choice of using the point of closest approach to the beam axis has a negligible effect. For cases where secondary vertexes are important, further development is needed to properly deal with the secondary vertex clustering and time reconstruction. 


The deterministic annealing includes a large class of algorithms with many tunable parameters. Only a few parameters are of interest for the application presented in this paper, that we discuss in more detail in subsequent sections. Parameters resulting in the choice of the energy function and the covariance matrix, the method chosen to assign tracks to clusters at the end of the cooling, and the choice of the temperatures at which we stop the annealing process are the most crucial ones.

To simplify the notation, the subscript $\textsc{ca}$ will be dropped in the following discussion. The energy between the track $i$ and the vertex prototype $k$ is defined as
\begin{equation}
\label{eq:Energy}
E_{ik} = \frac{(t_i - t_\textsc{v}^k)^2}{\sigma_{t,i}^2} + \frac{(z_i - z_\textsc{v}^k)^2}{\sigma_{z,i}^2}
\end{equation}
where $\{t_\textsc{v}, z_\textsc{v}\}$ is the prototype position and $\sigma_{z, i}$ ($\sigma_{t, i}$) is the uncertainty on the position (time) of closest approach for track $i$. Using this choice, the typical temperature ($T = 1/\beta$) scale of the vertexes is set to be of $O(1)$.
The track partition function is then defined as
\[
Z_i = \rho e^{-\beta \mu^2} + \sum_k e^{-\beta E_{ik}}
\]
where $\rho$ and $\mu$ are free parameters used to gauge the outlier rejection.
The parameter $\mu$ can be approximately interpreted as the number of standard deviations after which a track is called an outlier for a given vertex prototype. In this study, we fixed $\mu$ to 
$4$. The $\rho$ parameter is initially set to 0 and increased to 1 in small incremental steps at the end of the annealing loop. It is crucial to increment $\rho$ slowly in order to activate the outlier rejection quasi-adiabatically. 

To penalize particles with high impact parameter, each track is weighted according to:
\[
\displaystyle
w = \frac{1}{1+e^{\left(\frac{d_0}{\sigma_{d_0}}\right)^2 - S_0}}
\]
where $d_0$ is the reconstructed track impact parameter and $S_0$ is a parameter which 
determines when the impact parameter weighting becomes important. For 
our study, we set $S_0$ to be at $1$ standard deviation.
With the definition of $p(k,i) = p_k e^{-\beta E_{ik}} Z_i^{-1} $ and $p_k = \sum_i w_i p(k,i) / \sum_j w_j$, the vertex prototype time position is computed as
\[
t_\textsc{v}^k = \left(\sum_i \frac{p(k,i)w_i}{\sigma^2_{t,i}} t_i\right) \Big/ \left(\sum_i \frac{p(k,i)w_i}{\sigma^2_{t,i}}\right),
\]
and $z_\textsc{v}^k$ is computed similarly.

\begin{figure}[tbp]
\includegraphics[width=0.9\textwidth]{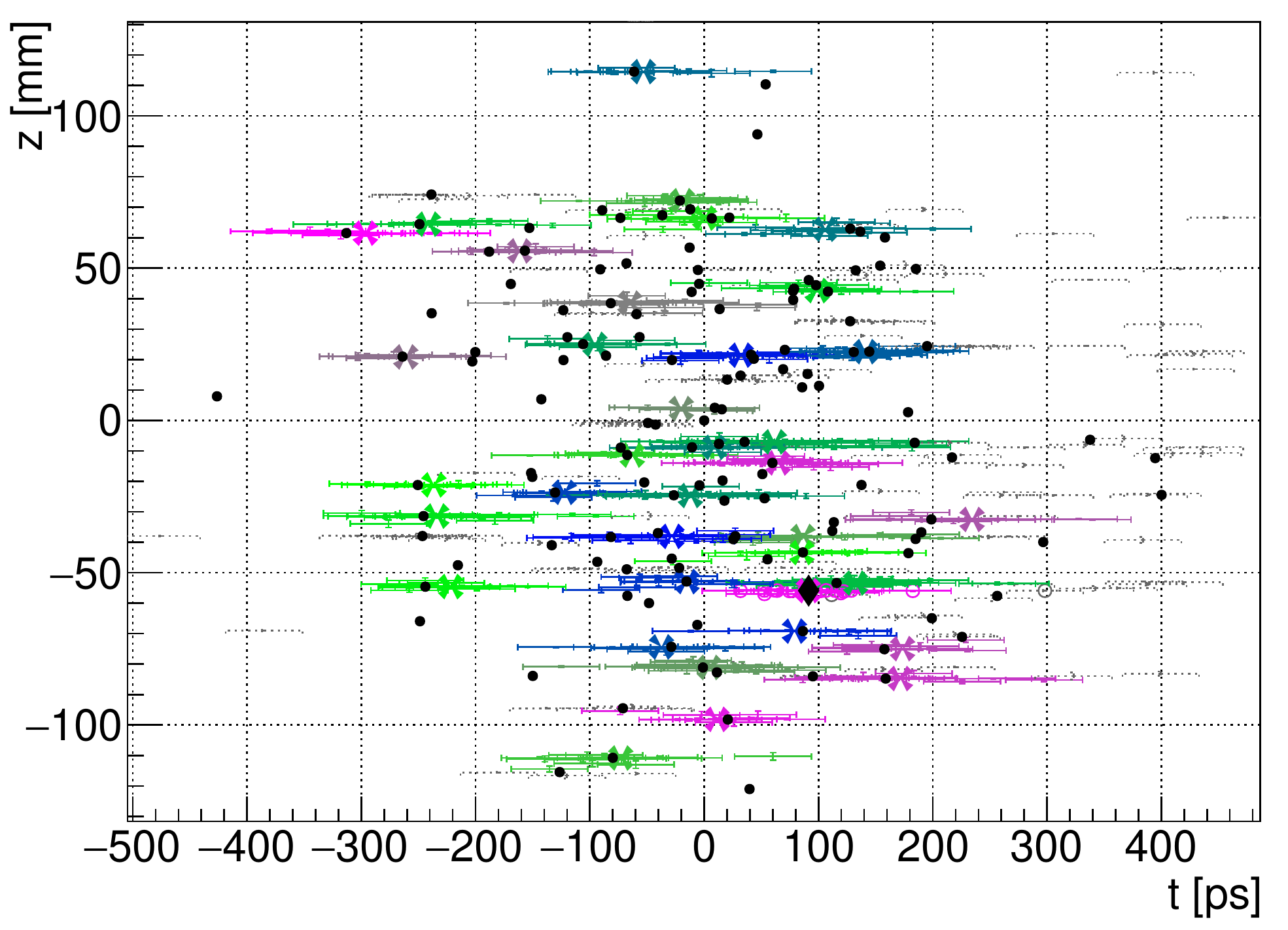}

\vspace*{-12cm}

\hspace*{9cm}\includegraphics[width=0.45\textwidth]{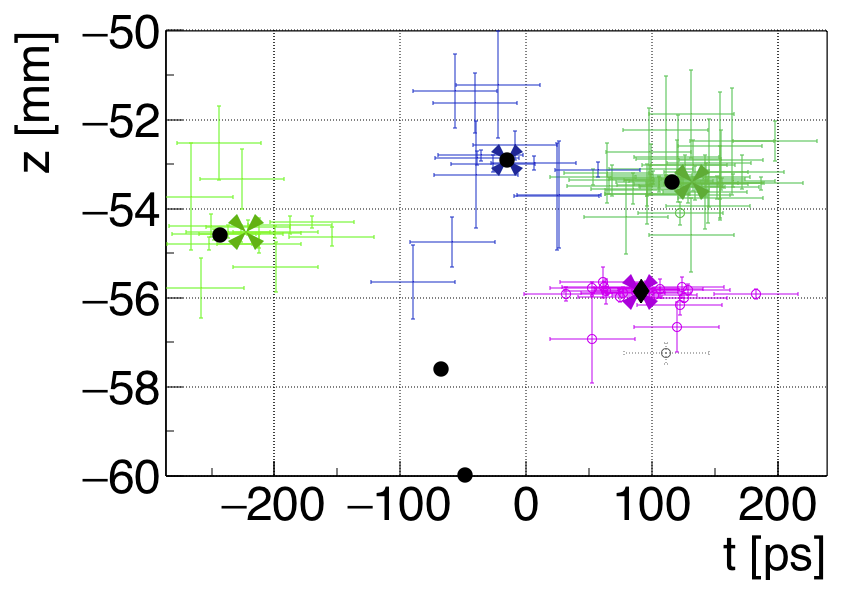}
\vspace{8cm}
\caption{The $t-z$ plane at the end of DA clustering for an example $gg \to \tilde{t}_1 
\bar{\tilde{t}}_1$ event with 140 pileup collisions. Representative points of each track are shown as error bars; the positions of clustered vertices are shown as crosses, and the true position of generated vertices are shown as black circles for pileup collisions, and black diamonds for the signal vertices. Tracks assigned to each vertex are shown using the same color. Tracks assigned to no vertex are shown in light gray.\newline
Top right corner: zoom of the region near the reconstructed signal vertex (purple).}
\label{fig:DA_end_shot}
\end{figure}

Further defining
\begin{align*}
p(i,k) &= \frac{w_i p(k,i)}{p_k}\\
w^{i,k}_{xy} &= \frac{p(i,k)}{\sigma_{x,i}\sigma_{y,i}}
\end{align*}
where both $x$ and $y$ can be $t$ or $z$, the vertex covariance matrix used has the form
\[
C^k_{xy} = \frac{\sum_i w^{i,k}_{xy} (x_i -x_\textsc{v}^k)(y_i -y_\textsc{v}^k)}{\sum_i w^{i,k}_{xy}}.
\]
The annealing cycle starts at $\beta=+\infty$, where the critical temperature for the only prototype ($\beta_0$) is computed. The system is immediately cooled down slightly above $\beta_0$.
The annealing loop, set as follows, is run until $\beta_{s} = 0.2$ is reached:
\begin{enumerate}
\item Prototype positions are updated until 
\[
\max_k \left[ 
\left(\frac{\Delta t_\textsc{v}^k}{\sigma_\textsc{v}^t}\right)^2 
+
\left(\frac{\Delta z_\textsc{v}^k}{\sigma_\textsc{v}^z}\right)^2 
\right] \leq 0.5 
\]
where $\Delta$ expresses the variation in the update cycle, and $\sigma_\textsc{v}^t = 10 \text{ ps}$ and $\sigma_\textsc{v}^t = 0.1 \text{ mm}$ are normalization factors chosen to represent the expected vertex resolution.
\item If two vertices are less than $2$~$\sigma$ apart, then they are consdiered not resolvable. Therefore, prototypes with a normalized distance smaller than 2 are merged and the cycle is updated.
\item The temperature is reduced by the cooling factor $f_C = 0.8$. We have observed that this parameter has small impact provided it is of order 1, which keeps the cooling process quasi-adiabatic.
\item Vertices below critical point are split into two along the maximum eigenvalue direction.
\end{enumerate}
At this point, a purging loop is run to remove prototypes with low probability or less than 2 tracks, for which that vertex is the closest. The procedure is repeated, cooling down the system until $\beta_{p} = 1$. Finally, a ultimate cooling is performed until $\beta_{M} = 1.5$ to sharpen the cluster border and tracks are assigned to the closer prototype.
Figure~\ref{fig:DA_end_shot}  shows a typical space-time configuration at the end of the DA clustering.

The cluster position gives a satisfactory estimate for the vertex position and in this study no further vertex fitting is performed. Tracks that are not assigned to a cluster are then potential candidates for a heavy charged particle. 

\subsection{Particle Identification}
\label{sec:PID}
A second Delphes module has been developed to identify potential HSCP candidates and, in general,  cluster particles which have not been classified because of the inconsistency of the $m_\pi$ hypothesis assumed at the beginning of the DA.
In practice, for each unmatched track with $|d_0|/\sigma_{d_0} < 3$, a two step procedure is followed. Standard Model particles are considered in the following order: pion, kaon, proton, electron and muon. For each mass hypothesis the $t_\textsc{CA}$ is recomputed and the compatibility with the vertices obtained from the DA is tested. The vertices are tested in the order of decreasing $\sum p_T^2$.  The track is assigned to the first vertex compatible to within $2\sigma$ and the mass is fixed to that given hypothesis. If no match is found for all of the above particle hypotheses, the track is passed to the second step. The vertex with the highest $\sum p_T^2$ that satisfies spatial compatibility is considered, and the mass which minimizes the closest approach distance from that vertex is estimated as the mass of that charged particle. 

The method developed has been verified to obtain reasonable performance in identifying heavy charged particles. Further details of the performance are
discussed in Section~\ref{sec:PIDPerformance}. Future improvements, including better criteria for the mass choice hypothesis or allowing the mass to be a free parameter in the DA clustering, may yield further improved results.

\subsection{Vertex Reconstruction and Particle Identification Performance}
\label{sec:PIDPerformance}
We evaluate the performance of the vertex reconstruction described above using a sample of simulated signal events with an injected $\tilde{t}_1$ mass of 500 GeV.

\begin{figure}
\centering
\includegraphics[width=0.8\textwidth]{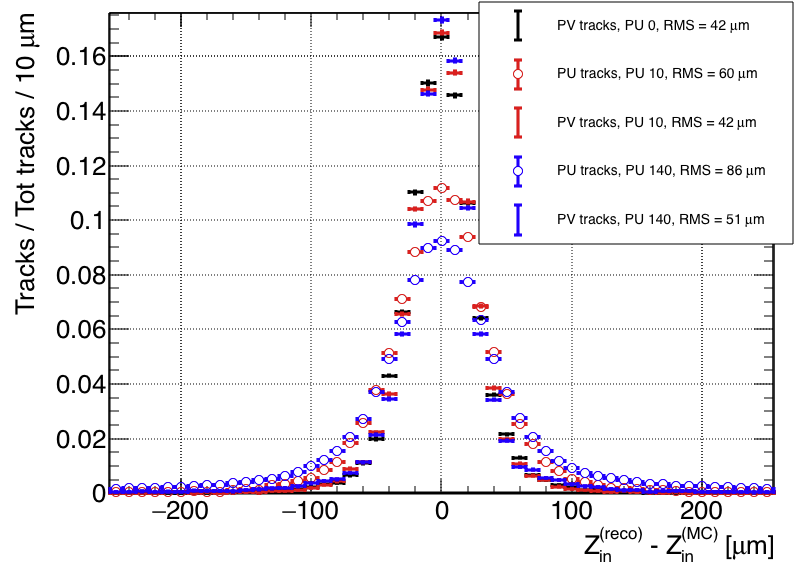}
\caption{The resolution on the track production point (vertex) $\hat{z}$ coordinate obtained from the vertex reconstruction procedure described in Section~\ref{sec:vertexing} from a sample of signal events is shown. Tracks from $p-p$ interaction which contains the generated signal process (PV) and from pile up interaction (PU) are show separately. Different colors correspond to the resolution for scenarios with different number of PU interaction per bunch crossing.}
\label{fig:Zin_res}
\end{figure}

\begin{figure}
\centering
\includegraphics[width=0.8\textwidth]{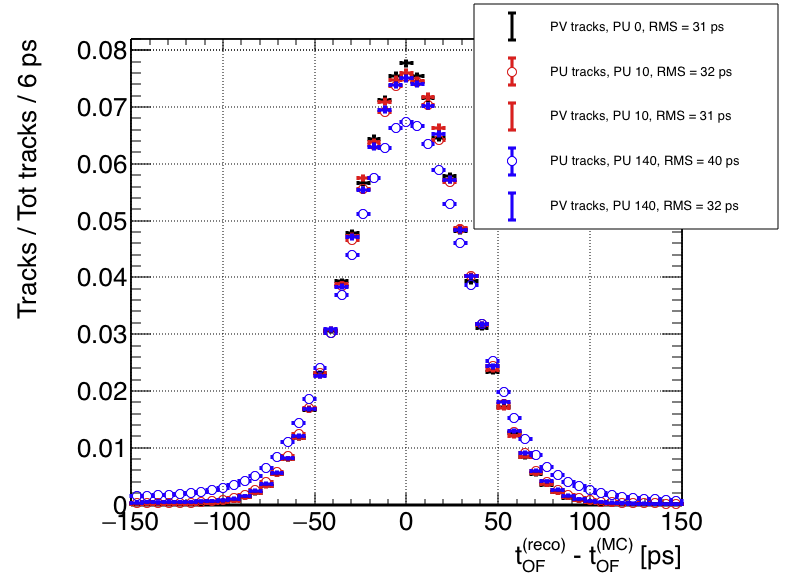}
\caption{The resolution on the TOF from a sample of signal events is shown. Tracks from $p-p$ interaction which contains the generated signal process (PV) and from pile up interaction (PU) are show separately. Different colors correspond to the resolution for scenarios with different number of PU interaction per bunch crossing.}
\label{fig:tof_res}
\end{figure}

In Figure~\ref{fig:Zin_res}, we show the difference between the true position simulated in the Monte Carlo generator and reconstructed production point $\hat{z}$ coordinate for each track. 
Tracks from the collision that produced the top squarks and R-Hadrons are labeled as PV and are shown separately from tracks resulting from pileup interactions. The performance for PV tracks is better compared to the pileup ones due to the higher number of tracks and larger transverse momentum. In Figure~\ref{fig:tof_res}, we show the analogous plot for the resolution on the TOF. The TOF resolution has a very small dependence on the amount of pileup and remains around $30$~ps even for a scenario with 140 pileup collisions per bunch crossing. 

\begin{figure}
\centering
\includegraphics[width=\textwidth]{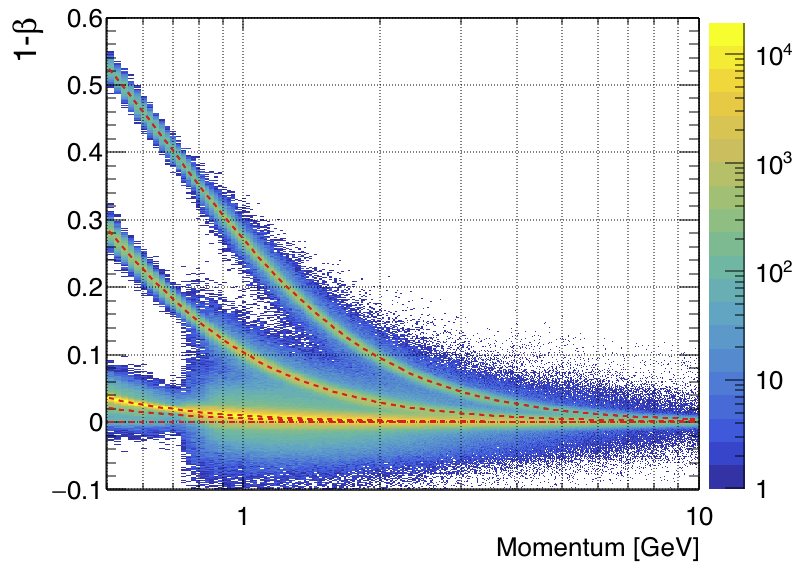}
\caption{The reconstructed velocity ($\beta$) versus the reconstructed momentum for tracks from both PV and PU interactions is shown for a scenario with 140 PU and 30 ps time-of-arrival resolution. Dashed red lines show the analytical relation for different masses: (from top to bottom) proton, kaon, pion, muon and electron mass.}
\label{fig:beta_VS_p}
\end{figure}

\begin{figure}[tbp]
\centering
\includegraphics[width=\textwidth]{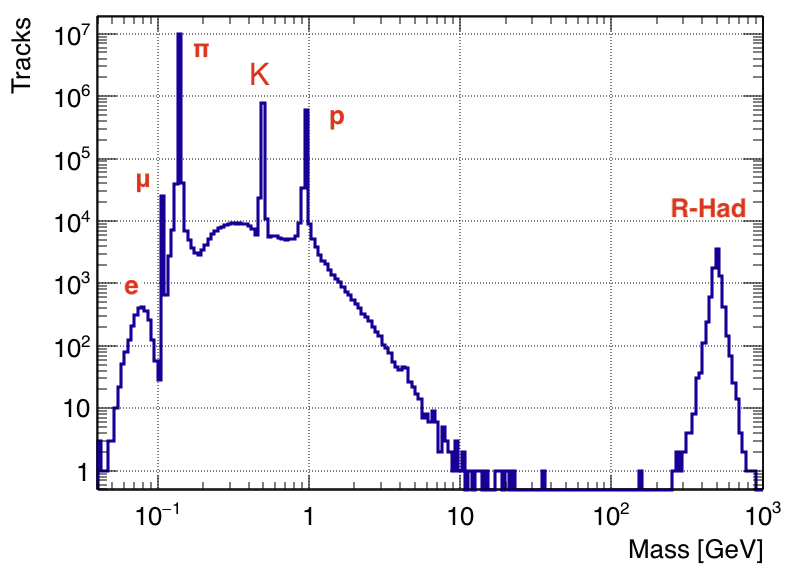}
\caption{Reconstructed mass spectrum from a sample of signal events  for a scenario with 140 PU and 30 ps time-of-arrival resolution. Tracks coming from the main collision and also those from pileup collisions are included in the plot.}
\label{fig:reco_mass_spectrum}
\end{figure}

To evaluate the particle identification power of the above resolution performance, we show a two dimensional histogram of the velocity versus the momentum (Figure~\ref{fig:beta_VS_p}) for all the tracks associated to a vertex in the same signal sample used in Figures~\ref{fig:Zin_res} and~\ref{fig:tof_res}. We can  observe separation between protons, kaons, and pions for momenta up to a few GeV, in agreement with the discussion from Section~\ref{sec:TOFPID}. The reconstructed R-Hadrons are very well separated from the SM particles and lie far outside the boundaries of the displayed plot.
Finally, Figure~\ref{fig:reco_mass_spectrum} shows the mass spectrum reconstructed using the TOF measurement as described in Section~\ref{sec:PID}. 

Peaks corresponding to the different particles can be clearly seen in the plot: electrons, muons, pions, kaons, protons and R-Hadrons, in order of increasing mass. For a $500$~GeV R-hadron we achieve a mass resolution of about $10\%$.

\section{Benchmark Search for Heavy Stable Charged Particles}
\label{sec:benchmark_analysis}
We perform a simple search analysis for heavy R-hadrons using the TOF measurement to illustrate the notable impact that a TOF detector would have at a proton-proton collider such as the HL-LHC. We compare the cross section limits for heavy stable charged particles (HSCPs) obtained with this search with existing limits from CMS and show that significant gains in sensitivity are possible with a new TOF detector. 

\subsection{Trigger}
\begin{figure}[tbp]
\centering
\includegraphics[width=\textwidth]{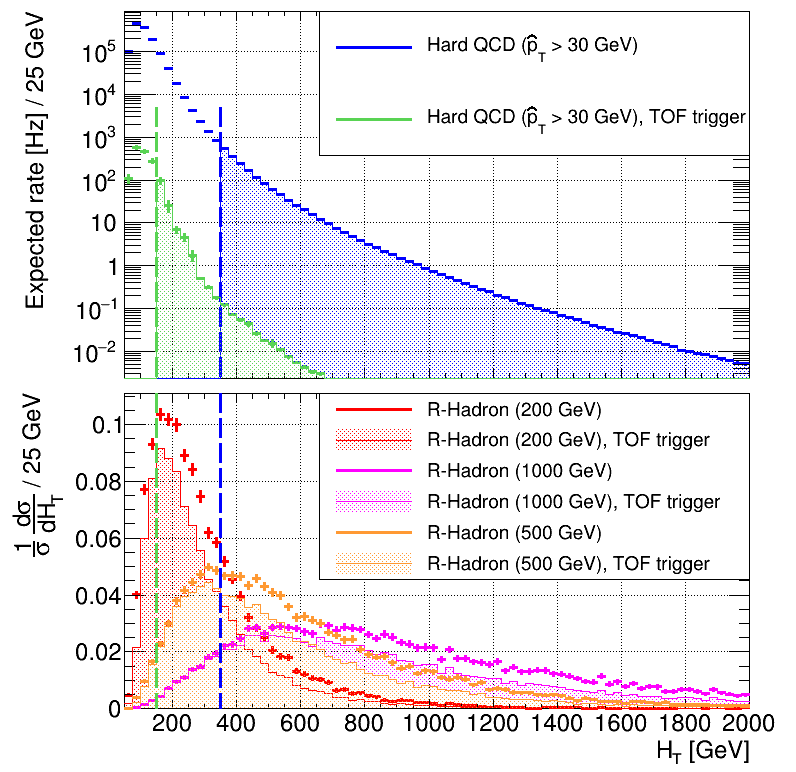}
\caption{Top: Expected rate of events for hard QCD interaction as a function of track $H_T$. The full distribution from all generated events (solid blue), and the distribution after the TOF track mass trigger requirements (solid green), are shown. Bottom: The differential distribution as a function of track $H_T$ for R-hadron production events where at least one R-hadron is within the detector acceptance are shown. The solid line shows the full distribution while the shaded area shows the distribution for events that pass the TOF track mass trigger. Different colors represent different stop mass samples. In both panels, the dashed lines show the track $H_{T}$ cut for the baseline scenario (blue), and the scenario with the TOF track mass trigger (green).}
\label{fig:SumPT}
\end{figure}

We consider two benchmark trigger scenarios for the R-hadron search. In the baseline scenario, we employ the proposed CMS L1 track trigger~\cite{CMSCollaboration:2015zni} to require large scalar sum of the transverse momentum ($H_T$) of all tracks associated with a particular collision vertex. Based on CMS studies~\cite{CMSCollaboration:2015zni}, the best estimate for a L1 track $H_T$ trigger with reasonable trigger rates, requires a threshold of $350$~GeV. In Figure~\ref{fig:SumPT}, we plot the track $H_{T}$ spectra for the background on the top panel and top squark signal on the bottom panel for a few different top squark mass points, along with the trigger threshold at $350$~GeV. For top squark masses above $500$~GeV, the track HT trigger will still retain more than $50\%$ of the signal. However for smaller top squark masses the signal efficiency decreases significantly. Therefore, we consider a second scenario with the added assumption that the TOF information becomes available in the L1 trigger. Seeded by tracks with $p_{T}>10$~GeV, and using a similar procedure as described in Section~\ref{sec:TOFRECO}, we require that the largest reconstructed track mass based on the TOF measurement is above $10$~GeV. Combining this track mass requirement with a less stringent track $H_{T}$ requirement of $H_{T} > 150$~GeV, allows us to reduce the background rate to a level similar to the rate of the more stringent $H_{T}>350$ trigger (below 150 Hz), while significantly improving the signal efficiency, for top squarks with mass below $200$~GeV, from about $20\%$ to above $80\%$. This specific scenario, as well as more generalized analyses of long-lived particle production~\cite{Liu:2018wte}  demonstrate potential of a TOF-based trigger in the upgraded CMS trigger system.

\subsection{Search strategy}
We consider a scenario where long-lived top squark pairs are produced and hadronize into stable R-hadrons in the detector volume. Events are split into two categories, one where R-hadrons from both top squarks are detected (the two R-hadron category), and one where an R-hadron from only one of the top squarks is detected (the single R-hadron category). Events are classified into the two R-hadron category if two R-hadron tracks are reconstructed with $p_{T}$ larger than $50$~GeV, and the relative difference in track mass is less than $10\%$. Otherwise, events are classified into the single R-hadron category if there is one R-hadron track with $p_T > 100$~GeV. 

\begin{figure}[tbp]
\centering
\includegraphics[width=0.9\textwidth]{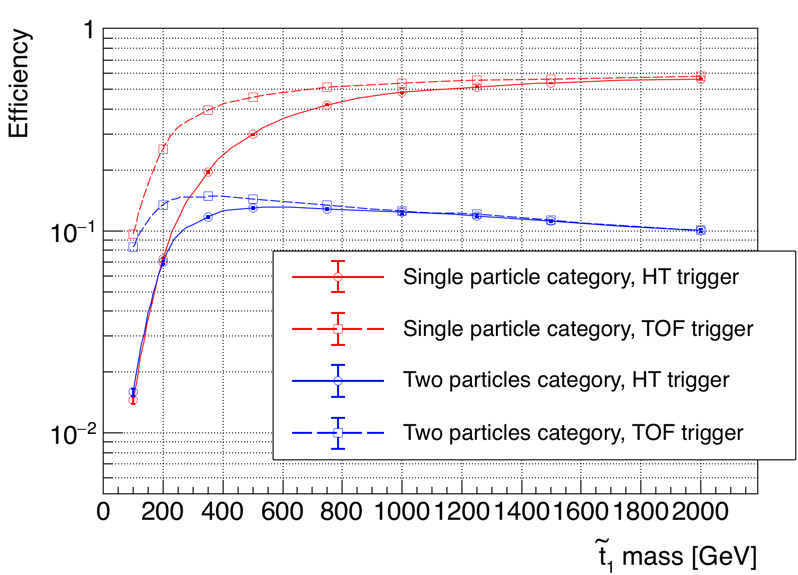}
\caption{Selection efficiency of simulated R-Hadron production events in the two different categories as a function of the generated $\tilde{t}_1$ mass. Efficiency is computed as the ratio of the number of events that pass trigger and category selection over the total number of generated events. Both $H_T$ (solid line) and TOF (dashed line) scenarios are presented.}
\label{fig:Efficiency_comparison}
\end{figure}
In Figure~\ref{fig:Efficiency_comparison} we show the acceptance times selection efficiency for signal events as a function of the generated top squark mass. This efficiency is dominated by the limited detector acceptance, particularly in the forward region. By comparing the efficiency with the baseline scenario that uses the $H_{T}$ trigger with the scenario where TOF is used in the trigger, we observe clearly that incorporating TOF measurements in the trigger system has a huge impact on enhancing the search sensitivity for top squark masses below $500$~GeV, improving the efficiency by up to an order of magnitude.

Finally, in each search category, we perform a fit to the reconstructed R-hadron track mass to extract the signal from the falling background spectrum. For the two R-hadron category, we define $\text{m}_{\text{h}}$ and $\text{m}_{\text{l}}$ as the larger and smaller mass of the two R-hadron tracks, and fit to the average of the two masses. In both categories, only events with $M > 50 \text{ GeV}$ are considered.

\subsection{Signal and Background Modeling}
\label{sec:modeling}
\begin{figure}[tbp]
\centering
\includegraphics[width=0.8\textwidth]{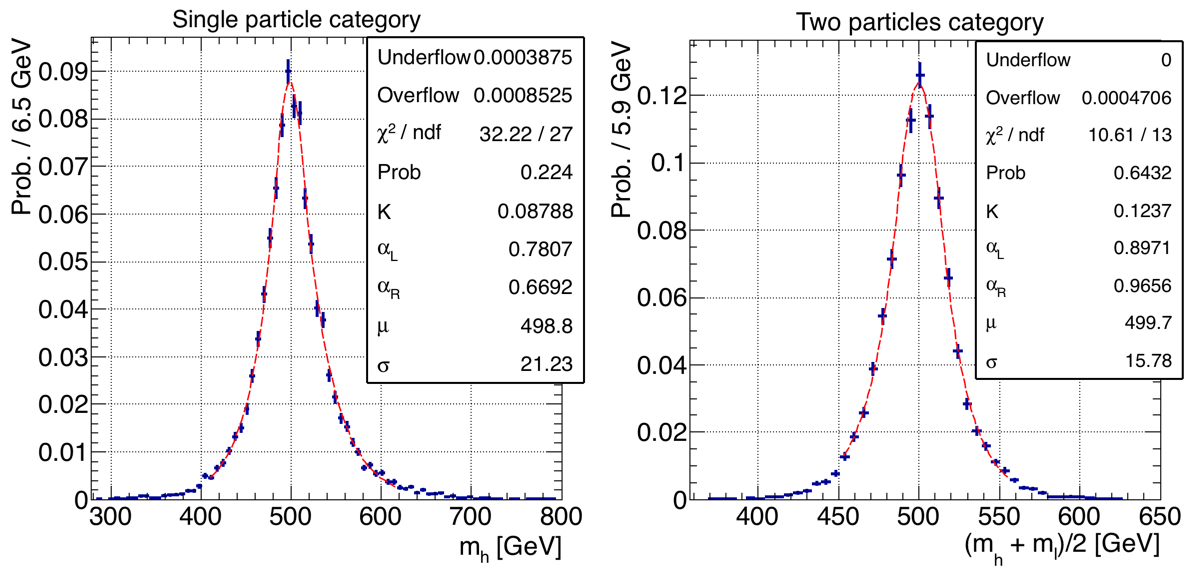}
\caption{Reconstructed mass spectrum in single particle (left) and two particles category (right) for simulated R-Hadron production event assuming $H_T$ trigger. The mass set in Pythia simulation for $\tilde{t}_1$ is 500 GeV. For both categories the dashed red line shows the best fit using a Gaussian function with asymmetrical exponential tails.}
\label{fig:RHad_mass500_spectrum}
\end{figure}

The signal mass shape is modeled, in both categories, as a Gaussian with asymmetric exponential tails. The model has a total of four shape parameters that are determined by a fit to the signal Monte Carlo sample: Gaussian mean ($\mu$), Gaussian width ($\sigma$) and the two exponential tail parameters ($\alpha_L$, $\alpha_R$). In Figure~\ref{fig:RHad_mass500_spectrum} we show an example of such a fit for a signal with top squark mass of $500$~GeV. For masses for which no simulated sample is available, the value of the Gaussian parameters and the exponential tail parameters are obtained by interpolating between mass points for which simulated samples were generated.

The signal is distinguished from the background through a mass reconstruction based on the time-of-flight of a charged particle. The background is primarily composed of events where the time-of-flight of a charged particle, produced through the standard model QCD multijet production process, has been instrumentally mis-measured. Mis-measurement of the vertex time and the arrival time of the charged particle both contribute, and are dominated by the effect of the time resolution of the time-of-arrival detector. These time mis-measurements result in an exponentially
decaying shape for the charged particle mass distribution for the dominant QCD multijet background. 

\begin{figure}[tbp]
\centering
\includegraphics[width=0.7\textwidth]{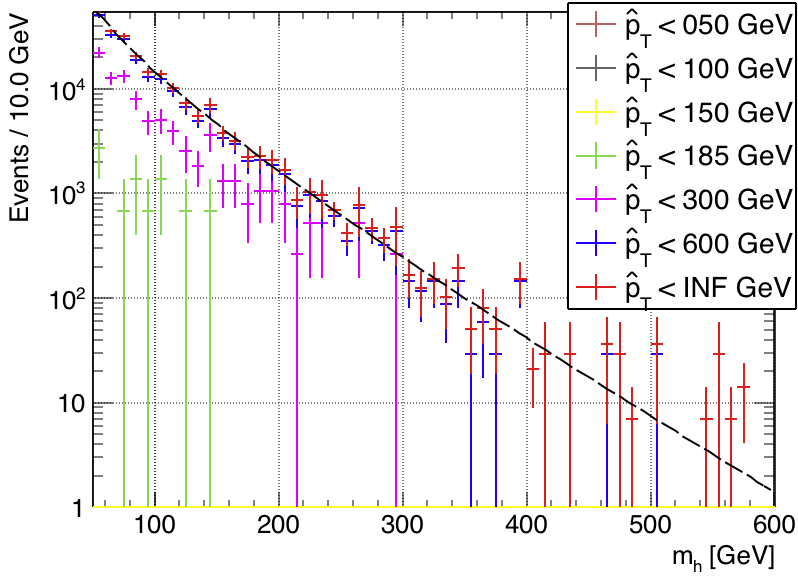}
\caption{Simulated distribution of the observable $M$ (corresponding to $m_h$ in this case) in hard QCD events passing the $H_T$ trigger and the single particle category requirements. The histograms of different colors represent the contribution from different $\hat{p}_t$ bins. The dashed black lines shows the best fit function used to model the background being $M^* = 67.0 \pm 0.3$ GeV the best fit parameter.\newline
The total number of events in the histogram is $2.24 \cdot 10^5$, normalized to a luminosity of $L=12.3 \text{ fb}^1$.
}
\label{fig:BkgMassSpectrumHTc1}
\end{figure}

We model the background mass spectrum by fitting an exponentially decaying analytic functional form to Monte Carlo samples of QCD multijet production. For the single R-hadron category, the following functional form is used: $P(M)dM \propto \frac{e^{-M/M^*}}{M} dM$, where $1/M^{*}$ is the exponential decay parameter extracted from the fit.  For the two R-hadron category the following functional form is used: $P(M)dM \propto e^{-M/M^*} dM$. In Figure~\ref{fig:BkgMassSpectrumHTc1}, we show an example of the background spectrum and the fitted functional form model 
for the single R-hadron category in the baseline track $H_{T}$ trigger scenario. The QCD multijet background Monte Carlo sample is generated in several bins of $\hat{p}_{T}$ in order to efficiently populate the full track mass spectrum.
Considering a luminosity of $L=12.3 \text{ fb}^1$, the number of events passing the cuts is $2.24 \cdot 10^5$ and the best fit parameter is $M^* = 67.0 \pm 0.3$ GeV.

\subsection{Results}
\begin{figure}[tbp]
\centering
\includegraphics[width=\textwidth]{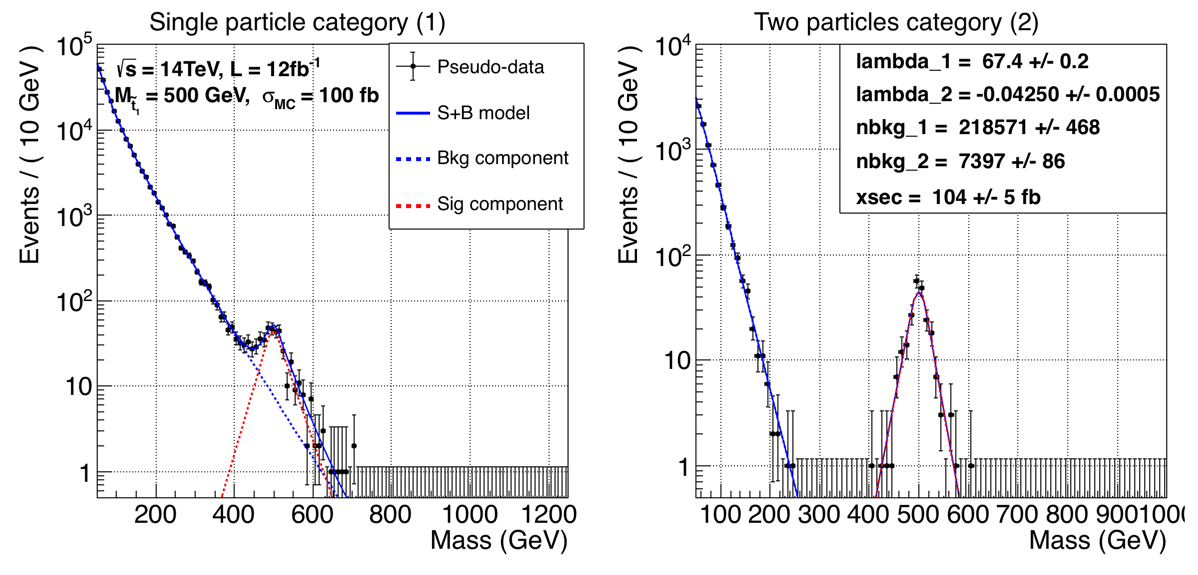}
\caption{Simulated mass spectrum (black dots) in the single particle (left) and two particles (right) categories for an integrated luminosity of 12.3~$fb^{-1}$ and a stop production cross section of $100$~fb. Signal (red) and background component (blue) are shown with dashed lines whereas the total fitting function is shown in solid blue. Best fit values of the fit free parameters are shown in the right panel.}
\label{fig:SimulFit_M500_L12}
\end{figure}

Based on the signal and background models derived in Section~\ref{sec:modeling}, we generate
pseudo-data for given integrated luminosity and signal cross sections. In Figure~\ref{fig:SimulFit_M500_L12}, we 
show an example of signal and background pseudo-data generated for $12$~$\mathrm{fb}^{-1}$ of integrated luminosity
for proton-proton collisions at a center of mass energy of $14$~TeV, and an assumed top squark mass of $500$~GeV and production cross section of $100$~fb. Fits of the signal and background in the single and two particle categories are performed simultaneously using the models described in Section~\ref{sec:modeling}.

\begin{figure}[tbp]
\centering
\includegraphics[width=\textwidth]{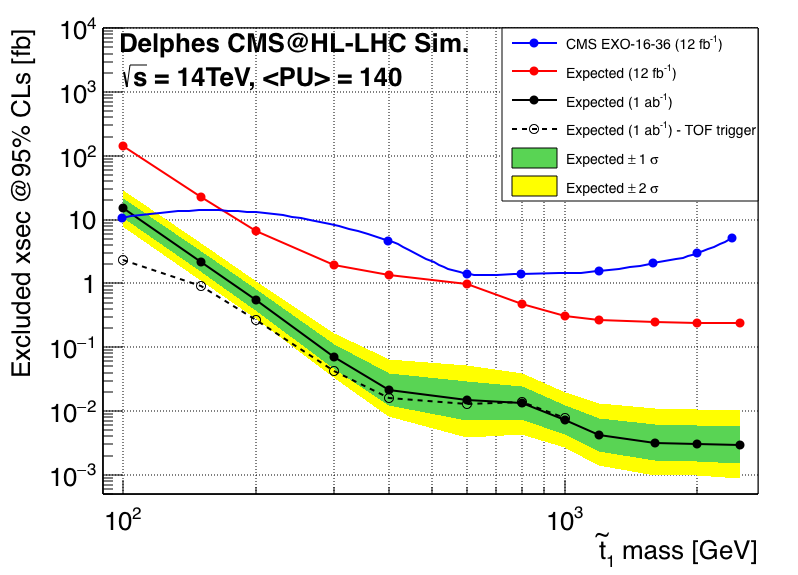}
\caption{Exclusion limits on the production of R-Hadrons at LHC. The improvement in sensitivity using TOF is discussed in the text.}
\label{fig:ExclusionLimits}
\end{figure}

Using the asymptotic approximation~\cite{cowan2011asymptotic} we derive the $95\%$ confidence level expected exclusion limits using the CLs method~\cite{Read:2002hq} for long-lived top squark production with lifetime sufficiently large that the top squark is stable over the full detector volume of the CMS detector. The expected limit for $12$~$\mathrm{fb}^{-1}$ of integrated luminosity is shown in Figure~\ref{fig:ExclusionLimits} and compared to the best existing CMS limits~\cite{CMS-PAS-EXO-16-036}. We show that the sensitivity of this analysis using the TOF measurement is better than limits that do not use timing information for top squark masses above about $170$~GeV. The expected limit for our analysis improves more sharply as the top squark mass increases because the larger mass results in slower velocities and increased time delay, which our analysis is sensitive to, while the sensitivity of the best existing CMS limits are less dependent on the top squark mass. Therefore, the improvement over the existing CMS limits is generally enhanced for larger top squark masses and ranges from a factor of $5$ to $10$. We also compare the expected limit at $1$~$\mathrm{ab}^{-1}$ of integrated luminosity between the baseline scenario using the $H_{T}$ trigger and the scenario where we employ TOF measurements in the trigger, and we observe that at top squark mass below $200$~GeV the TOF-based trigger improves the sensitivity by a factor of $2$ to $5$. 

\section{Summary}
\label{sec:summary}
In view of the future proposed timing capabilities of the LHC experiments during HL-HLC, we studied the impact of a time of flight measurement in performing particle identification. 
We computed the analytical formula for the expected performance and, given the foreseen timing resolution, we estimated the particle identification potential to be significant within SM particles up to a transverse momentum of few GeV. Similarly, we computed the expected peak width in the measurement of a heavy (500 GeV) stable particle mass with the TOF and we found it to be of the order of 10\% of the mass.

Using \textsc{Pythia} to generate the processes and Delphes to perform a fast simulation of the upgraded CMS detector for HL-LHC, we proposed an approach to perform a TOF measurement with  minimal assumptions. Deploying a deterministic annealing to reconstruct vertices, we achieve a $\hat{z}$ resolution of about 50 (80)~$\mu$m for PV (PU) tracks for TOF resolution of about 30 ps. These resolutions are demonstrated to be sufficient to identify both SM and BSM particles.

Using long-lived top squark pair production as a benchmark example, we have demonstrated that significant sensitivity gains in searches for long-lived particles can be made with the aid of a dedicated time-of-flight detector. We demonstrate how such a detector would enable four-dimensional vertex reconstruction and the identification of charged particles through its time-of-flight measurement. Mass resonances with good resolution can be reconstructed solely on the basis of the particle track and its time-of-flight, and can significantly enhance the rich program of searches for heavy stable charged particles. We demonstrate for our benchmark example an improvement in sensitivity of a factor of $5$ to $10$ for top squark masses above $300$~GeV. Finally, we show that if the time-of-flight measurement could be utilized in the trigger system, an additional sensitivity improvement
of a factor of $2$ to $5$ could be realized for top squark masses below $200$~GeV. This result, along with concurrent complementary studies~\cite{Liu:2018wte}, provide good  motivation for further work on the design and realization of a time-of-flight based trigger for long-lived particles.

\acknowledgments

This work has been supported by funding from California Institute of Technology High Energy Physics under Contract DE-SC0011925 with the United States Department of Energy. We thank  Cliff Cheung, Lindsey Grey, Artur Apresyan, Maurizio Pierini, Josh Bendavid, Jonathan Lewis,  the CMS MTD group, and the CMS Caltech group for very useful discussions and suggestions. MS is grateful to Henry Frisch for many years discussions on precision timing systems and PID of heavy exotic particles.

\bibliographystyle{JHEP}
\bibliography{PID.bib}

\end{document}